%% file: dalembert.tex
\documentclass[12pt]{article}
\usepackage{makeidx}
\usepackage{epsfig}
\usepackage{float}
\usepackage{amscd}
\usepackage{amsmath}
\usepackage{amssymb}
\usepackage{color}
\usepackage{fancyheadings}
\usepackage{fancyvrb}
\usepackage{algorithm2e}
\usepackage{algpseudocode}

\usepackage{listings}
\usepackage{xcolor}

\definecolor{codegreen}{rgb}{0,0.6,0}
\definecolor{codegray}{rgb}{0.5,0.5,0.5}
\definecolor{codepurple}{rgb}{0.58,0,0.82}
\definecolor{backcolour}{rgb}{0.95,0.95,0.92}

\lstdefinestyle{mystyle}{
    backgroundcolor=\color{backcolour},   
    commentstyle=\color{codegreen},
    keywordstyle=\color{magenta},
    numberstyle=\tiny\color{codegray},
    stringstyle=\color{codepurple},
    basicstyle=\ttfamily\footnotesize,
    breakatwhitespace=false,         
    breaklines=true,                 
    captionpos=b,                    
    keepspaces=true,                 
    numbers=left,                    
    numbersep=5pt,                  
    showspaces=false,                
    showstringspaces=false,
    showtabs=false,                  
    tabsize=2
}

\usepackage{slashbox}
\usepackage{caption}
\usepackage{subcaption}

\input machip

\date{}

\makeindex
\usepackage{makeidx}

\begin{document}
\title{Resolution of d'Alembert's Paradox \\
Using Slip Boundary Conditions: The Effect of the Friction Parameter on the Drag Coefficient}
\author{Ingeborg Gjerde, Simula Research Library \\
L. Ridgway Scott, University of Chicago}
 \def\thepage{}\maketitle\pagenumbering{arabic}
\centerline{\today}

\begin{abstract}
d'Alembert's paradox is the contradictory observation that for incompressible and inviscid (potential) fluid flow, there is no drag force experienced by a body moving with constant velocity relative to the fluid. This paradox can be straightforwardly resolved by considering Navier's slip boundary condition. Potential flow around a cylinder then solves the Navier--Stokes equations using friction parameter $\beta=-2\nu$. This negative friction parameter can be interpreted physically as the fluid being accelerated by the cylinder wall. This explains the lack of drag.

In this paper, we introduce the Navier slip boundary condition and show that choosing the friction parameter positive resolves d'Alembert's paradox. We then further examine the effect of the friction parameter $\beta$ on the drag coefficient. In particular, we show that for large $\beta$ the drag coefficient corresponds well with experimental values. Moreover, we provide numerical evidence that the Newton continuation method (moving from small to large Reynold's numbers) requires fewer iterations to succeed. Thus the slip boundary condition is advantageous also from a computational perspective.
\end{abstract}

d'Alembert is famous for his observation that potential flow has zero drag. In a SIAM Review paper in 1981 \cite{stewartson1981d}, matched asymptotic expansions
(including the Triple Deck) were used to analyze this paradox. We return here to this subject from a different point of view, one that would have
been difficult in 1981. We consider the paradox by approximating solutions of the Navier-Stokes equations with a slip boundary condition,
using recently developed numerical techniques \cite{lrsBIBiy}, and of course
current computer platforms. In \cite{stewartson1981d}, an attempt was made to understand the paradox by
seeking a limiting solution for large Reynolds numbers. Similarly, in \cite{hoffman2010resolution}, an explanation of the paradox was attempted
using the Euler equations as a base. Here we show, by numerical computation, that the paradox is easily resolved
at modest Reynolds numbers, provided suitable boundary conditions are applied.

 In a recent paper \cite{lrsBIBiw} it was observed that potential flow around a bluff body is an exact solution of the Navier-Stokes equations when a relation holds between the kinematic viscosity $\nu$ and the Navier friction coefficient $\beta$. For a cylinder, the relationship is that $\beta=-2\nu$. This explains one defect of potential flow, namely that it corresponds to a slip condition with negative friction, which would require an active wall. We show that d'Alembert's paradox is resolved by restricting this friction parameter to be positive (or zero). The drag force on the cylinder is then non-zero, and increases together with the friction parameter and viscosity. 
 
 Further, we examine in more detail the effect the friction parameter has on the drag coefficient. Notably, we find that using the slip boundary condition leads to fewer iterations required for a Newton continuation method moving from small to large Reynold's numbers. Thus, the the slip boundary condition allows us to reliably compute drag coefficients for large Reynold's numbers. For large enough friction parameters $\beta$, we get solutions that approximately satisfy a no-slip boundary condition on the cylinder. The drag coefficients computed for these flows agree closely with experimental values. Finally, we find that the drag coefficient is not sensitive to changes in $\beta$ (for $\beta$ large enough). Thus, there is no need to identify the "correct" friction parameter for flows that should approximately satisfy a no-slip boundary condition. 
 
 The drag coefficient functional can be split into two parts, the pressure drag and the viscous drag. The latter involves the strain related to the flow field, multiplied by the viscosity (proportional to one over the Reynolds number). We observe that the strain goes to infinity like the square root of the Reynolds number. Fortunately, the viscous drag coefficient involves the stress (viscosity times strain), and this actually goes to zero as the Reynolds number increases. Thus, while drag is easily computable, other functionals may not behave reliably as the Reynolds number is increased.

We are concerned with base flows that are steady.
For unsteady flows, see \cite{ref:refvaldragcylinder}.
The steady flow we exhibit at Reynolds number 1000 is well beyond the
point where unsteady flows (Karman vortex streets) \cite[Figure 14.16]{PantonRonaldL2013IF} can be observed.
Thus our computation of steady flows at that Reynolds number is of
independent interest. 

The article will proceed as follows. To begin, we introduce in Section \ref{sec:modeqs} the model equations we consider, including the Navier slip boundary condition. Next, we give in Section \ref{sec:potflow} the potential flow solution, and show that it is an exact solution of the Navier--Stokes equations with slip boundary conditions if $\beta=-2\nu$.
In Section \ref{sec:varform},  we give the variational formulation of the model equations and show how to discretize them using $\mathbb{P}^2$--$\mathbb{P}^1$ (Taylor--Hood) elements.
In Section \ref{sec:daladrag}, we investigate the drag coefficient using slip boundary conditions. In Section \ref{sec:drag}, we give the definition of the viscous and pressure drag coefficients and study their dependence on $\beta$. In Section \ref{sec:expmeass}, we discuss how these agree with experimental values. In section \ref{sec:slipimact}, we describe the effect of $\beta$ on computational efficiency, and show the slip boundary condition leads to fewer iterations needed for a Newton continuation method. Finally, we comment in Section \ref{sec:asymptions} on how our computations raise some questions regarding the validity of using the viscosity as a small parameter for asymptotic expansions.

\omitit{
A significant question arises regarding the effect of instabilities on the
drag coefficient \cite{ref:newTheoryOFlight}.
It could well be that the drag coefficient related to laminar flow would be
significantly altered by perturbations of the flow.
The method of Reynolds and Orr \cite{ref:SerrinStabilityReOrr,lrsBIBiw} 
predicts a basis for unstable modes, and we examine the contribution to drag
for the most dominant unstable modes.
}

\section{Model equations}
\label{sec:modeqs}

\begin{figure}
\begin{subfigure}{0.5\textwidth}
\includegraphics[width=2.9in,angle=0]{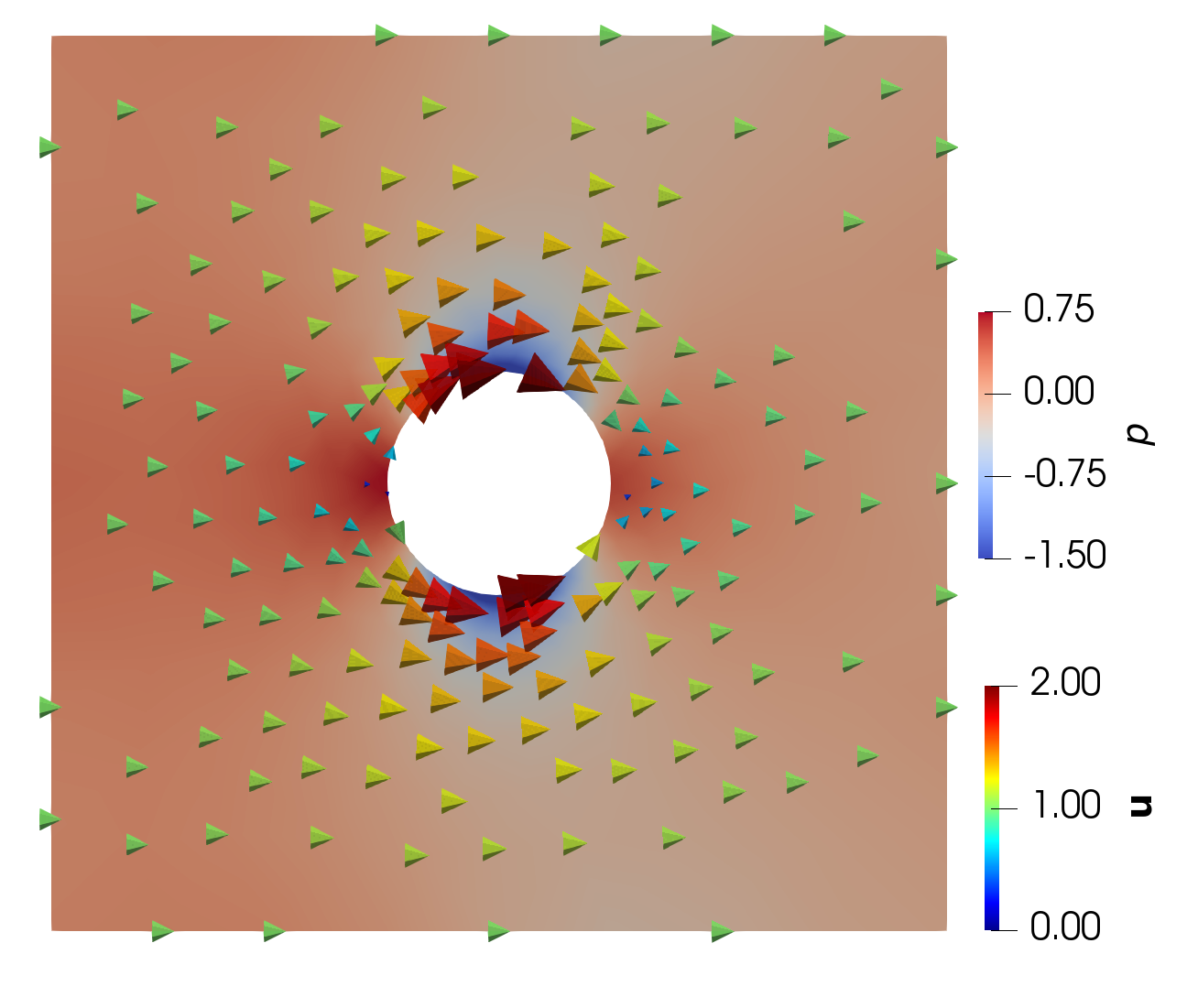}
\caption{$\beta=-2$}
\label{fig:betaminustwo}
\end{subfigure}
\begin{subfigure}{0.5\textwidth}
\includegraphics[width=2.9in,angle=0]{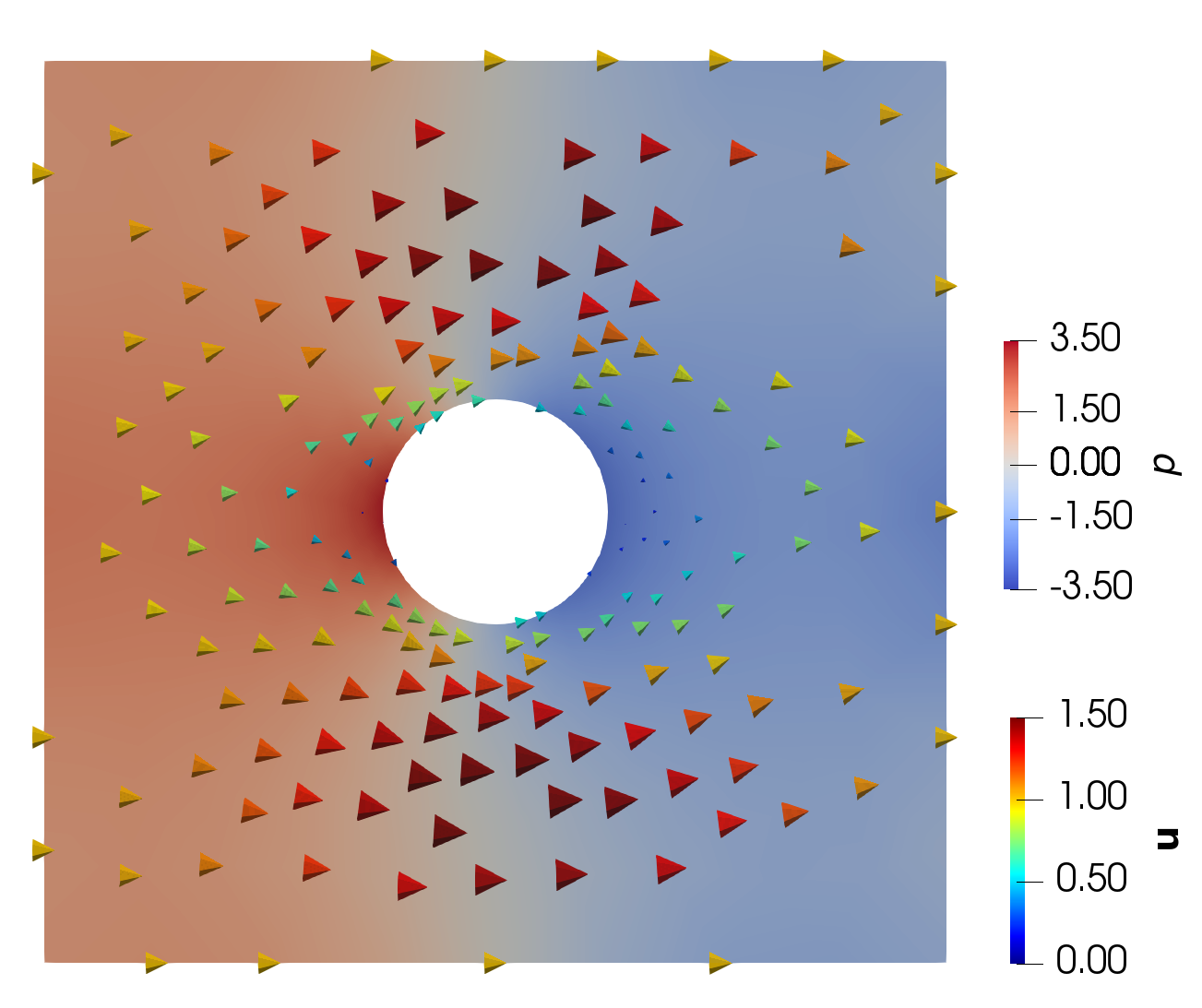}
\caption{$\beta=2$}
\label{fig:betaplustwo}
\end{subfigure}
\caption{Solution of the Navier-Stokes equations 
(\ref{eqn:firstnavst},\ref{eqn:bceesnavst},\ref{eqn:navierslip}) for $\nu=1$ with 
(a) $\beta=-2$ and (b) $\beta=2$.
The computation was done in a box of dimensions $8\times 8$, and the cylinder of radius 1 is 
centered in the box. Dirichlet conditions on the boundary of the box were given as $\uu=(1,0)$.}
\label{fig:betatwo}
\end{figure}

Suppose that $(\uu,p)$ is a solution of the steady Navier-Stokes equations
in a domain $\Omega$ containing an obstacle with boundary $\Gamma\subset\bo$:
\begin{equation} \label{eqn:firstnavst} 
\begin{split}
 -\nu\Delta \uu &+ \,\uu\cdot\nabla \uu 
         + \nabla p = \bfz\;\hbox{in}\;\Omega,\\
&\sdiv\uu =0\;\hbox{in}\;\Omega,
\end{split}
\end{equation}
where $\nu$ is a nondimensional parameter related to the kinematic viscosity,
together with boundary conditions
\begin{equation} \label{eqn:bceesnavst} 
\uu=\gbc\;\hbox{on}\;\partial\Omega\backslash\Gamma,\quad
\uu\cdot\nn =0\;\hbox{on}\;\Gamma,
\end{equation}
with $\nn$ being the outward pointing unit normal of $\Omega$, and Navier's slip condition \cite{ref:SlipNewtonianReviewexpts,lrsBIBiy}
linking tangential velocity and the shear stress on $\Gamma$:
\begin{equation} \label{eqn:navierslip} 
\beta \, \uu\cdot\btau_k = -\nu \, \nn^t(\nabla\uu+\nabla\uu^t)\btau_k,\quad k=1,2,
\end{equation}
where $\btau_k$ are orthogonal tangent vectors and $\beta$ a friction coefficient.
We assume that $\gbc=\bfz$ on $\Gamma$ for simplicity.

The sign in \eqref{eqn:navierslip} requires some explanation.
First of all, note that the sign of the tangent vectors does not matter, as they
appear on both sides of the equation.
However, the direction of the normal matters.
In one direction, the quantity on the right-hand side of \eqref{eqn:navierslip} gives
the shear force on the fluid caused by the cylinder, whereas the opposite direction gives
the (equal and opposite) shear force on the cylinder caused by the fluid.
We will see that it makes mathematical sense to have $\beta$ of either sign,
but typically $\beta\geq 0$ unless the cylinder has an active wall.

In Figures \ref{fig:betaminustwo} and \ref{fig:betaplustwo} we show two solutions of  \eqref{eqn:firstnavst}-\eqref{eqn:navierslip} for $\beta=-2$ and $\beta=2$, respectively. 
For $\beta=-2$, the solution is approximately equal to potential flow, with the flow being accelerated around the cylinder. This retains fore-aft symmetry of the pressure. Consequently, the drag is negligible; using \eqref{eqn:dragfullmb} the drag coefficient is calculated to $C_D = -0.08 \approx 0$. For $\beta=2$, the flow is significantly de-accelerated around the cylinder. This gives rise to a fore-aft asymmetry for the pressure and a consequent drag-coefficient of $C_D=9.4$.

In the next sections we will investigate further how the drag coefficient changes with $\beta$. From this point, we keep $\beta\geq 0$ as this makes sense physically. For the sake of comparison, we will also consider the Navier-Stokes equations \eqref{eqn:firstnavst}-\eqref{eqn:bceesnavst} with a no-slip boundary condition on $\Gamma$: 
\begin{align}
\uu = \bfz \text{ on } \Gamma. \label{eqn:no-slip}
\end{align}

\subsection{Potential flow as a solution of the Navier--Stokes equations with slip}
\label{sec:potflow}

Potential flow around a cylinder of radius 1 and aligned with the 
$z$-axis is given \cite{lrsBIBiw} by $\uu=\nabla\phi$ where
\begin{equation} \label{eqn:fiformpotf} 
\phi(x,y,z)=
\phi(r,\theta,z)=\Big(r+\frac{1}{r}\Big)\cos\theta=x+\frac{x}{x^2+y^2},
\end{equation}
The solution satisfies $\uu\cdot\nn=0$ on $\Gamma$, and the
velocity components are then given by
\begin{equation} \label{eqn:prevelcalcun} 
u_{x}(x,y,z) =1-\frac{x^2-y^2}{\big(x^2+y^2\big)^2}, \qquad
u_{y}(x,y,z)=\frac{-2xy}{\big(x^2+y^2\big)^2}.
\end{equation}
\omitit{
By simple calculus, we can compute $\nabla\uu$:
\begin{equation} \label{eqn:rgadhyun} 
r^6\,\nabla\uu=
\big(x^2+y^2\big)^3\,\nabla\uu=
\begin{pmatrix} 
{2x^3- 6x y^2} & {-2y^3 +6x^2 y}  \\
{-2y^3+6x^2y}  & {-2x^3+6xy^2}  \\
\end{pmatrix} .
\end{equation}
}
It can be verified \cite{lrsBIBiw}  that potential flow is a solution of Navier-Stokes 
\omitit{
stems from the vector-calculus identity
\begin{equation} \label{eqn:vecalcidan} 
\uu\cdot\nabla\uu+\uu\times(\nabla\times\uu)=\half\nabla|\uu|^2.
\end{equation}
For $\uu=\nabla\phi$, $\nabla\times\uu=\bfz$.
Since $\Delta\phi=0$, $\Delta\uu=\nabla\Delta\phi=\bfz$.
Thus with $\uu=\nabla\phi$,
$$
-\nu\Delta\uu+\uu\cdot\nabla\uu=\frac{1}{2}\nabla|\uu|^2,
$$
and we have a solution of 
}
\eqref{eqn:firstnavst} for any $\nu$ with
\begin{equation} \label{eqn:peearridan} 
p=-\frac{1}{2}|\uu|^2=-\frac{1}{2}|\nabla\phi|^2.
\end{equation}
Similarly, it can be verified  \cite{lrsBIBiw} that
\omitit{
Next, let us compute $\beta$ so that the slip boundary condition \eqref{eqn:navierslip} is satisfied. To this end, take the normal and tangent vectors on the cylinder to be
$$
\nn=-(x,y),\qquad \btau=(y,-x),
$$
From \eqref{eqn:prevelcalcun}, we have that the left side of the Navier slip condition equals
\begin{equation} \label{eqn:appcaldot} 
\uu\cdot\btau= (2y^2,-2xy)\cdot(y,-x)=2y^3+2x^2y=2y
\end{equation}
on the cylinder. 

As $\nabla \uu$ is symmetric, we have $\nabla \uu + \nabla \uu^t = 2\nabla \uu$. Using \eqref{eqn:rgadhyun} we find
\begin{equation} \label{eqn:youtauxn} 
\begin{split}
(\nabla\uu)\btau &=
\begin{pmatrix} 
{2x^3- 6x y^2} & {-2y^3 +6x^2 y}  \\
{-2y^3+6x^2y}  & {-2x^3+6xy^2}  
\end{pmatrix} 
\begin{pmatrix} y \\ -x  \end{pmatrix} \\
& =
\begin{pmatrix} y(2x^3- 6x y^2)-x(-2y^3 +6x^2 y) \\ 
y(-2y^3+6x^2y)-x(-2x^3+6xy^2) \end{pmatrix}  \\
& = \begin{pmatrix} -4x^3 y - 4x y^3 \\ -2y^4+ 2x^4 \end{pmatrix} 
=\begin{pmatrix} -4xy  \\ -2y^4+ 2x^4 \end{pmatrix} 
\end{split}
\end{equation}
on the cylinder. Combining this we have
\begin{equation} \label{eqn:normauxn} 
\nn^t(\nabla\uu + \nabla\uu^t)\btau = 2\nn^t(\nabla\uu)\btau =-2(x,y)\begin{pmatrix} -4xy  \\ -2y^4+ 2x^4 \end{pmatrix} 
                       =4y\big(2x^2+y^4- x^4\big).
\end{equation}
But
$$
2x^2+y^4- x^4=2x^2+(1-x^2)^2 - x^4=1
$$
and thus the right sde of the Navier slip condition equals simply $4y$ on the cylinder. Therefore 
}
\eqref{eqn:navierslip} is satisfied for $\beta=-2\nu$.

Thus potential flow for the cylinder provides an exact solution of Navier-Stokes
for any Reynolds number, and without any boundary layer, for $\beta=-2 \nu$.
This may be viewed as nonphysical, as it represents an active boundary condition:
the tangential stress increases the tangential velocity. For $\beta=-2\nu$, the force from this active boundary condition perfectly balances the drag force, and the cylinder experiences no drag.

\subsection{Variational formulation and discretization}
\label{sec:varform}

Define the space $V$ by
\begin{equation} \label{eqn:veespascdf}
V=\set{\vv\in H^1(\Omega)^d}{\sdiv\vv =0\;\hbox{in}\;\Omega,\quad \vv=\bfz\;\hbox{on}\;
\partial\Omega\backslash\Gamma,\quad \vv\cdot\nn =0\;\hbox{on}\;\Gamma}.
\end{equation}
We also define the shear stress
$$
\dv=\nabla\vv+\nabla\vv^t 
$$ 
and the following inner products
$$
(\uu,\vv)=\intox{\uu\cdot\vv},\qquad (\du,\dv)=\intox{\du : \dv}.
$$
where $\uu,\vv$ are vectors. 

Consider the tangent space $T$ to $\Gamma$ and the projection $P_T$ onto the tangent space. In \cite{lrsBIBiy}, we show that the variational formulation of \eqref{eqn:firstnavst}--\eqref{eqn:navierslip} is:

Find $\uu \in V$ such that:
\begin{equation} \label{eqn:newvarfst} 
\frac{\nu}{2}(\du,\dv)+(\uu\cdot\nabla\uu,\vv) 
+\oint_{\textcolor{blue}{\Gamma}} \beta(P_T \vv)\cdot (P_T \uu) \,ds=0
\end{equation}
for all $\vv\in V$, since $(p,\sdiv\vv)=0$ for $\vv\in V$.

In order to discretize \eqref{eqn:firstnavst}-\eqref{eqn:navierslip}, we consider a Nitsche method for imposing the Navier slip condition \eqref{eqn:navierslip}. Any solution $\uu \in V$ also solves the following variational formulation (for any constants $h,\gamma \in \mathbb{R}$) \cite[Lemma 2.1]{lrsBIBiy}:

Find $\uu \in H^1(\Omega)$, $p \in L^2(\Omega)$ and $\lambda \in \mathbb{R}$ such that
\begin{equation}
A((\uu,p,\lambda),(\vv,q,\sigma)) + (\uu \cdot \nabla \uu, \vv)= 0 \label{eqn:nitsche}
\end{equation}
for all $(\vv,q,\sigma)\in H^1(\Omega)$, $q \in L^2(\Omega)$ and $\sigma \in \mathbb{R}$, where
\begin{gather}
\begin{aligned}\label{eqn:goodischtstok}
A((\uu,p,\lambda),(\vv,q,\sigma)) &=a(\uu,\vv)+b(\uu,q) +b(\vv,p) + \int_{\Omega} {\rho \, q+\sigma \, p}\,d\xx.
\end{aligned}
\end{gather}
Here, the forms $a$ and $b$ are given as
\begin{equation}\label{eqn:goodpartsstok}
\begin{split}
a(\uu,\vv)&=
\frac{\nu}{2}(\du,\dv)  
+ \oint_{\Gamma}  \beta  (\uu\cdot\btau) (\vv\cdot\btau) \,ds \\
&- \oint_{\Gamma} \nu\,\nn^t\du\nn\, (\nn\cdot\vv) \,ds 
- \oint_{\Gamma} \nu\,\nn^t\dv\nn \,(\nn\cdot\uu) \,ds \\
& +\gamma \oint_{\Gamma} h^{-1}  (\uu\cdot\nn) (\vv\cdot\nn) \,ds ,
\end{split}
\end{equation}
and
\begin{equation}\label{eqn:goodpbeesstok}
b(\vv,q) = - (q,\sdiv\vv) + \oint_{\Gamma} q\,(\nn\cdot\vv) \,ds 
\end{equation}

The variational formulation \eqref{eqn:nitsche} can then be discretized directly using Taylor-Hood $\mathbb{P}^2$--$\mathbb{P}^1$ elements for the flux and pressure, respectively. Taking $h$ as the mesh size, this formulation is stable given that the Nitsche parameter $\gamma>0$ is large enough. This method is validated in \cite{lrsBIBiy}, using, e.~g., potential flow as an analytic solution and computing the corresponding error rates.

\section{d'Alembert and drag}
\label{sec:daladrag}

d'Alembert is famous \cite{LandauLifshitz} for noticing that the drag on a cylinder 
(or other body) is zero for inviscid potential flow.

Before we define the drag coefficient, let us recall how \eqref{eqn:firstnavst} is derived. 
First, let us recall the dimensional, time-dependent Navier--Stokes equations. We write $\hat\xx=L\xx$,
$\hat\uu(\hat\xx,t)=U\uu(L^{-1}\hat\xx,t)$,
and $\hat p(\hat\xx,t)=\rho U^2 p(L^{-1}\hat\xx,t)$,
where $L$ is the radius of the cylinder (taken to be 1 here).

The dimensional Navier--Stokes equations read \cite{LandauLifshitz} 
$$
\hat\uu_t-\frac{\mu}{\rho}\Delta\hat\uu+\hat\uu\cdot\nabla\hat\uu+\frac{1}{\rho}\nabla\hat p=0 ,
$$
where $\mu$ is the dynamic viscosity and $\rho$ is the density.
The dimensions of each term in this equation are force divided by mass, which
are the units of acceleration, length divided by time squared.
Computing derivatives of the relationships between the functions and their
hatted versions, we get (with $\hat\nu=\mu/\rho$)
$$
U\uu_t-\frac{\hat\nu U}{L^2}\Delta\uu+\frac{U^2}{L}\uu\cdot\nabla\uu
+\frac{U^2}{L}\nabla p=0.
$$
Multiplying this by $L/U^2$, we get
$$
\frac{L}{U}\uu_t-\frac{\hat\nu}{L U} \Delta\uu+\uu\cdot\nabla\uu +\nabla p=0.
$$
The quantity $\tau=L/U$ is a time unit.
The quantity $R=UL/\hat\nu$ is dimensionless and is called the Reynolds number.
Thus we get
$$
\tau\uu_t-\frac{1}{R} \Delta\uu+\uu\cdot\nabla\uu +\nabla p=0.
$$
In our computations, we have taken $L=2$ (cylinder radius 1), so
$\nu=2/R$ ($R=2/\nu$) in \eqref{eqn:firstnavst} and subsequent relations.

The force $\ff_\bchi$ on a body with surface $\widehat\Gamma$ 
in a direction $\hat\bchi$ is given by 
\begin{equation} \label{eqn:dragbionpa} 
\begin{split}
\ff_\bchi&=\oint_{\widehat\Gamma} \mu \big(\dhu\hat\bchi\big)\cdot\hat\nn 
                 - \hat p \, \hat\nn\cdot\hat\bchi\,d\hat s
 =\oint_{\Gamma} \frac{\rho\hat\nu U}{L} \big(\du\bchi\big)\cdot\nn 
                 - \rho U^2 p \, \nn\cdot\bchi \,ds \\
&= \rho U^2 \oint_{\Gamma} \frac{\hat\nu}{LU} \big(\du\bchi\big)\cdot\nn 
                 -  p \, \nn\cdot\bchi\, ds \\
&= \rho U^2 \oint_{\Gamma} \nu \big(\du\bchi\big)\cdot\nn 
                 -  p \, \nn\cdot\bchi\, ds,
\end{split}
\end{equation}
since $\hat\nn=L^{-1}\nn$, $\hat\bchi=L^{-1}\bchi$, and $d\hat s=L^2 ds$.
For a cylinder, this is zero by symmetry for potential flow if $\bchi=(1,0)$
is the flow direction. Thus, for potential flow, the cylinder will not experience lift nor drag.

\subsection{Drag coefficient for different friction parameters}
\label{sec:drag}
In this section, we study the dependence of the drag and lift coefficients on the friction parameter $\beta$. The drag and lift coefficients are often defined via
\begin{equation} \label{eqn:seedeeraginy} 
C_\chi=\frac{2}{A} \frac{\ff_\bchi}{\rho U^2} =\frac{2}{A}
 \oint_\Gamma \nu(\du \bchi)\cdot\nn - p\nn\cdot\bchi  \,d\xx,
\end{equation}
where $A$ is the cross-sectional area of the obstacle represented by $\Gamma$.
These coefficients depend on $\nu$, $\beta$, and $\Gamma$.
For a cylinder of radius 1, $A=2$.
The drag coefficient $C_D$ for a cylinder correpsonds to $\bchi=(1,0)^t$ on $\Gamma$,
and the lift coefficient $C_L$ for a cylinder correpsonds to $\bchi=(0,1)^t$ on $\Gamma$.

The two parts of the drag in \eqref{eqn:dragbionpa} have different names.
The form drag (a.k.a.~pressure drag) coefficient $C_P$ for a cylinder is
\begin{equation} \label{eqn:dragformpa} 
  C_P=- \oint_{\Gamma}  p \, \nn\cdot\bchi\, ds.
\end{equation}
The skin friction drag (a.k.a.~viscous drag) coefficient $C_V$ for a cylinder is
\begin{equation} \label{eqn:dragformmb} 
C_V=\oint_{\Gamma} \nu \big(\du\bchi\big)\cdot\nn\, ds.
\end{equation}
Thus the full drag coefficient for the cylinder is given by 
\begin{equation} \label{eqn:dragfullmb} 
C_D=C_P+C_V.
\end{equation}

\begin{figure}
\centering
R=1 \includegraphics[width=0.99\textwidth]{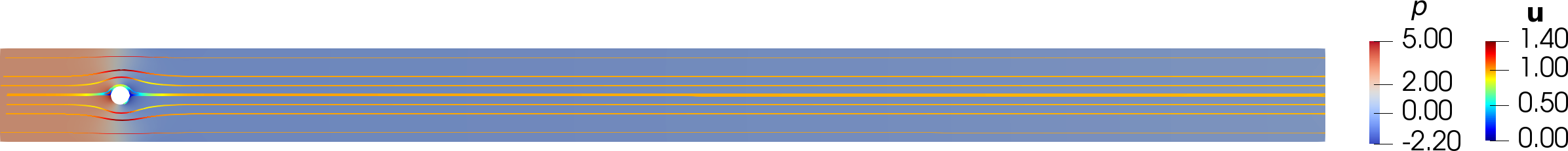}
R=10 \includegraphics[width=0.99\textwidth]{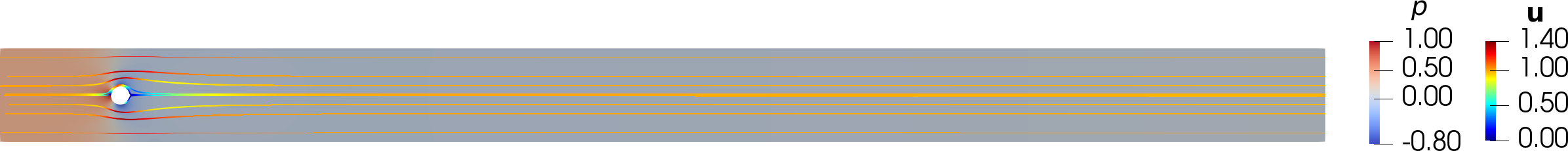}
R=100 \includegraphics[width=0.99\textwidth]{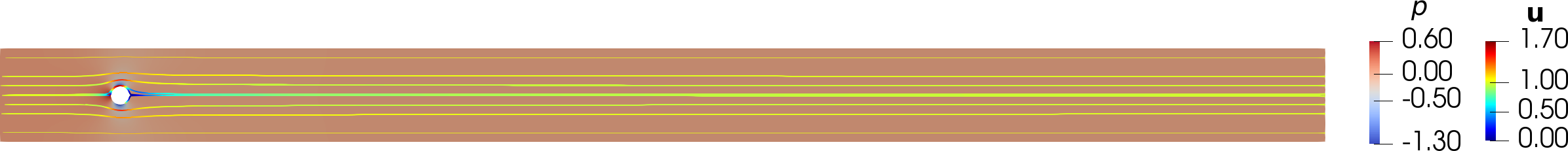}
R=1000 \includegraphics[width=0.99\textwidth]{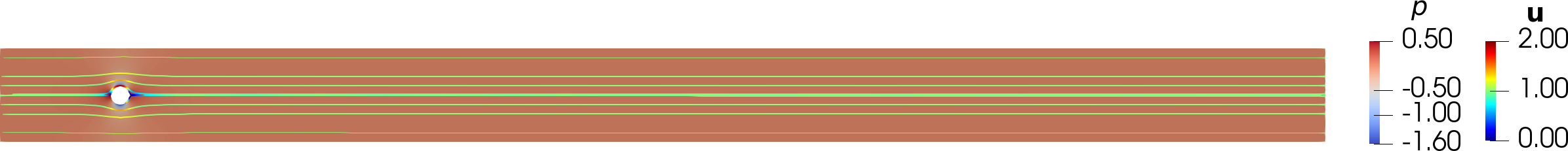}
\caption{Velocity streamlines and pressure for the solution of the Navier--Stokes equations with slip boundary conditions and friction parameter $\beta=0$ and different Reynolds numbers $R$. (Top) Results for $R=1$; significant fore-aft asymmetry in the pressure leads to a large pressure drag. This pressure difference decreases as the Reynolds number increases. (Bottom) Results for $R=1000$; the pressure solution is approximately symmetric and we expect a small pressure drag.
Computations were done on the domain $\Omega$ defined in \eqref{eqn:longdomain}.}
\label{fig:betazero}
\end{figure}
\begin{figure}
\centering
R=1 \includegraphics[width=0.99\textwidth]{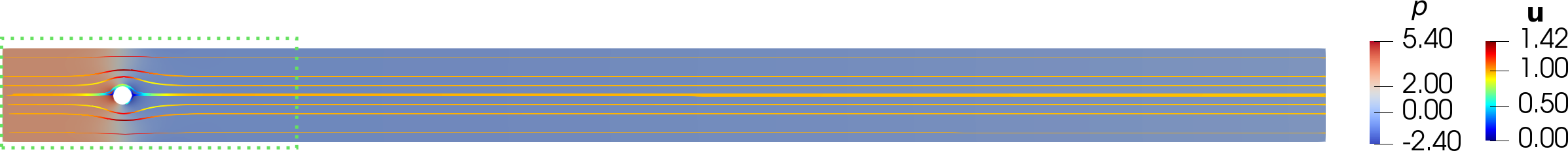}
R=10 \includegraphics[width=0.99\textwidth]{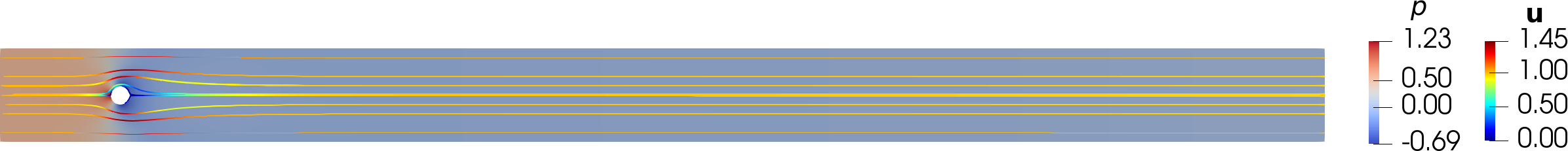}
R=100 \includegraphics[width=0.99\textwidth]{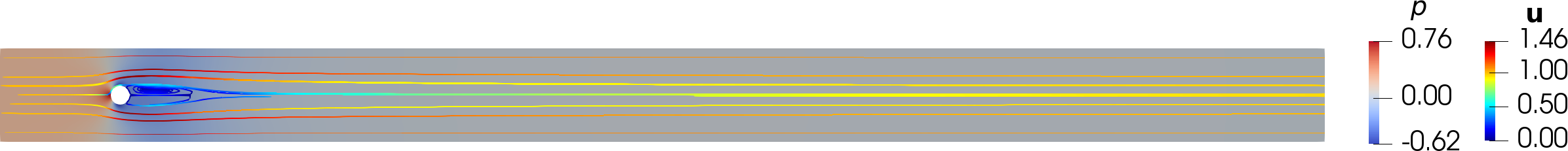}
R=1000 \includegraphics[width=0.99\textwidth]{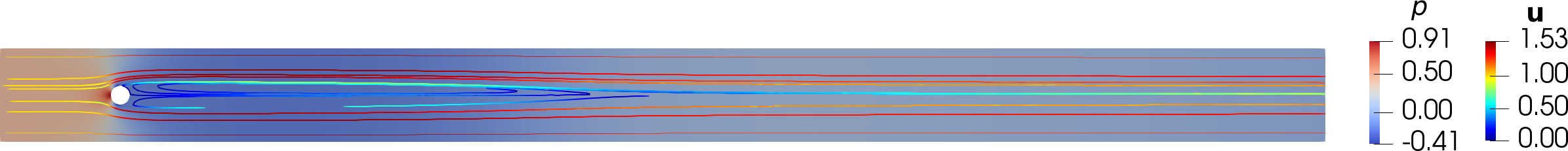}
\caption{Velocity streamlines and pressure for the solution of the Navier--Stokes equations with slip boundary conditions and friction parameter $\beta=1$ and different Reynolds numbers $R$. (Top) Results for $R=1$; significant fore-aft asymmetry in the pressure around $\Gamma$ leads to a large pressure drag. As the Reynolds number increases, the pressure drop from fore to aft gets smaller and the wake behind the cylinder increases. (Bottom) Results for $R=1000$; there is a large wake behind the cylinder. Comparing the results with $R=100$, we see that the pressure drop from fore to aft has stabilized; consequently, we expect the pressure drag coefficient to stabilize as well.
Computations were done on the domain $\Omega$ defined in \eqref{eqn:longdomain}.}
\label{fig:betaone}
\end{figure}
\begin{figure}
\centering
R=1 \includegraphics[width=0.99\textwidth]{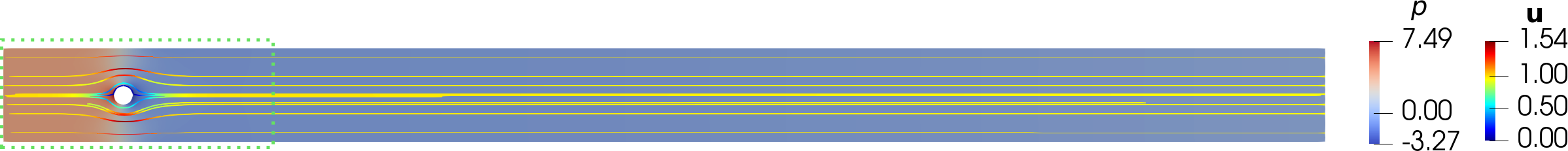} 
R=10 \includegraphics[width=0.99\textwidth]{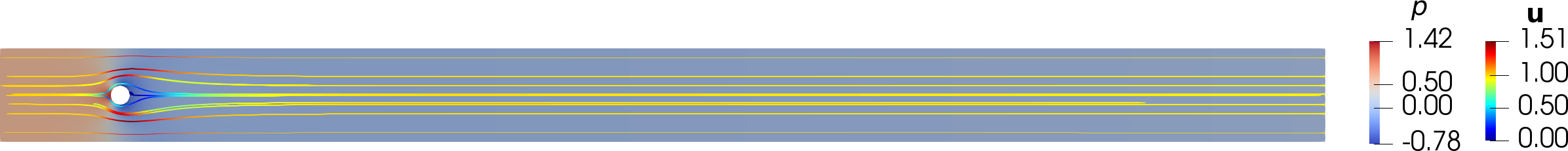} 
R=100 \includegraphics[width=0.99\textwidth]{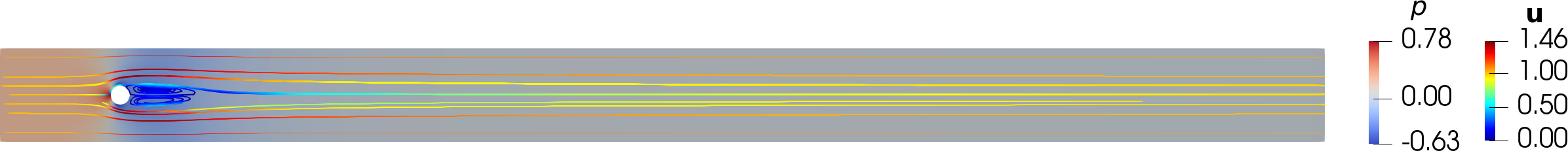} 
R=1000 \includegraphics[width=0.99\textwidth]{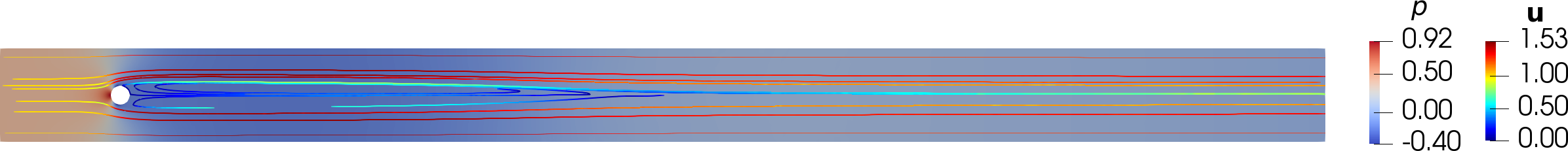}
\caption{Velocity streamlines and pressure for the solution of the Navier--Stokes equations with slip boundary conditions and friction parameter $\beta=100$ and different Reynolds numbers $R$. As the Reynolds number increases, we see that the difference from the solution with $\beta=1$ becomes negligible.Computations were done on the domain $\Omega$ defined in \eqref{eqn:longdomain}.}
\label{fig:betaK}
\end{figure}

In Figures \ref{fig:betazero}, \ref{fig:betaone} and \ref{fig:betaK} we show the solutions of Navier-Stokes equations with slip boundary conditions past a cylinder of radius 1 enclosed in a rectangular domain
\begin{align}\label{eqn:longdomain}
\Omega = \{(x,y) : -12.8 < x < 128, -5 < y < 5 \} \setminus \{ (x,y) \in \mathbb{R}^2 : x^2+y^2 \leq 1^2 \}.
\end{align}
We compute the solution for different friction parameters and Reynold's numbers; Figure \ref{fig:betazero} shows the solution for $\beta=0$; Figure \ref{fig:betaone} shows the solution for $\beta=1$; and Figure \ref{fig:betaK} shows the solution for $\beta=100$. For each of these, we compute the solution for Reynold's number $R=1, 10, 100$ and $1000$. For all simulations, we employ a Dirichlet boundary condition on the sides of the box, setting $\uu=(1,0)$ on $\partial \Omega \setminus \Gamma$. To solve the Navier--Stokes equations with slip boundary conditions in $\Omega$, we use the Nitsche method \eqref{eqn:nitsche} with 
Taylor--Hood elements \cite{lrsBIBih}, that is,
continuous piecewise quadratic $\mathbb{P}^2$-elements for the velocity and continuous piecewise linear $\mathbb{P}^1$-elements for the pressure. The Nitsche parameter is set to $\gamma=25$.


In Tables \ref{tab:fulldrag} and \ref{tab:viscdrag} we give the full and viscous drag, respectively, on the cylinder $\Gamma$. We see from Table \ref{tab:fulldrag} that for $\beta \geq 1$ the drag coefficients are close to the expected 
experimental results in the range of $R$ values considered \cite{ref:cylendardralow,ref:cylendardrag}.
We also see that there is little variation in drag coefficient for $\beta\geq 1$. Thus a small bit of friction resolves d'Alembert's Paradox. 

Note that for Reynold's number $R=1$, we needed to choose a wider domain in order to get the values indicated. In the narrower domain \eqref{eqn:longdomain}, we got the drag values 11.346, 12.232, 15.303 and  17.396 for $\beta=0,1,10$ and 100, respectively. For the no-slip boundary condition, we got the drag value 17.864 on the narrower domain. Thus, the domain geometry does affect the drag coefficients.

Comparing Figure \ref{fig:betazero} with the relevant drag coefficients in Tables \ref{tab:fulldrag} and \ref{tab:viscdrag} (i.e., the values given in the first row), we see that the total drag corresponds well with the degree of fore-aft asymmetry of the pressure. For $R=1$, there is asymmetry in the pressure and this gives rise to substantial drag coefficient as is seen in the first row in Table \ref{tab:fulldrag} .For $R=10$ this asymmetry gets smaller and for $R\geq 100$, the pressure is roughly the same before and after the cylinder, causing the pressure drag coefficient to be small.


\begin{table}
\begin{center}
\begin{tabular}{|c|*{4}{c|}}
\hline
\backslashbox[2em]{$\beta$}{$R$} & 1 & 10 & 100 & 1000 \\ 
\hline
 0  &  7.575 &  1.626 &  0.293 &  0.032 \\
1  & 8.012 &  2.379 &  1.227 &  1.090 \\ 
10  & 9.412 &  2.717 &  1.256 &  1.093 \\ 
100  & 10.259 &  2.763 &  1.258 &  1.093 \\ \hline
no-slip  & 10.401 &  2.784 &  1.257 &  1.052 \\\hline
experimental & 11.517 & 2.961 & 1.339 & 1.015  \\
\hline
\end{tabular}
\end{center}
\vspace{-6mm}
\caption{Drag coefficients $C_D$ defined in \eqref{eqn:dragfullmb}
for the cylinder for different friction coefficients
$\beta$ and Reynolds numbers $R$.
Experimental results taken from \cite[Figure 14.15, page 351]{PantonRonaldL2013IF}.
Computations were done on the domain $\Omega$ defined in \eqref{eqn:longdomain},
except in the case $R=1$ where the domain  from the cylinder axis to the
end is 64 and the width is 20.}
\label{tab:fulldrag} 
\end{table}

\begin{table}
\begin{center}
\begin{tabular}{|c|*{4}{c|}}
\hline
\backslashbox[2em]{$\beta$}{$R$} & 1 & 10 & 100 & 1000 \\ 
\hline
0  & -5.2e-03 &  -3.7e-04 &  -5.0e-05 &  4.6e-07 \\ 
1  & -4.2e-03 &  -7.0e-05 &  2.3e-05 &  -1.1e-05 \\ 
10  & -7.9e-04 &  9.9e-05 &  2.9e-05 &  -1.3e-05 \\ 
100  & 1.5e-03 &  1.3e-04 &  3.0e-05 &  -1.4e-05 \\ \hline
no-slip  & 4.3e-04 &  5.6e-05 &  1.2e-05 &  -4.9e-06 \\ 
\hline
\end{tabular}
\end{center}
\vspace{-6mm}
\caption{Viscous drag coefficients $C_V$ defined in \eqref{eqn:dragformmb} 
for the cylinder for different friction coefficients
$\beta$ and Reynolds numbers $R$.
Computations were done on the domain $\Omega$ defined in \eqref{eqn:longdomain}.}
\label{tab:viscdrag} 
\end{table}


In Figure \ref{fig:betaone} we see the pressure and velocities for $\beta=1$ and $R\in \{ 1, 10, 100, 1000 \}$. The asymmetry of the pressure field before and after the cylinder is evident. However, the pressure difference fore and aft of the cylinder gets smaller as the Reynolds number increases. From $R=100$ to $R=1000$, there is only a small change in the pressure drop fore and aft of the cylinder. This is consistent with the drag coefficients reported in Table \ref{tab:fulldrag}, which get (i) smaller as $R$ gets larger and (ii) seemingly stabilize to $C_D \approx 1$ as the Reynolds number gets large. The wake behind the cylinder grows with the Reynolds number. This impacts the viscous drag, as we can see in the second row in Table \ref{tab:viscdrag}. The viscous drag is, however, negligible compared to the full drag. Thus, we see no clear dependence of the full drag on the wake behind the cylinder.

\begin{figure}
\centering
\begin{subfigure}{0.9\textwidth}
\includegraphics[width=0.7\textwidth]{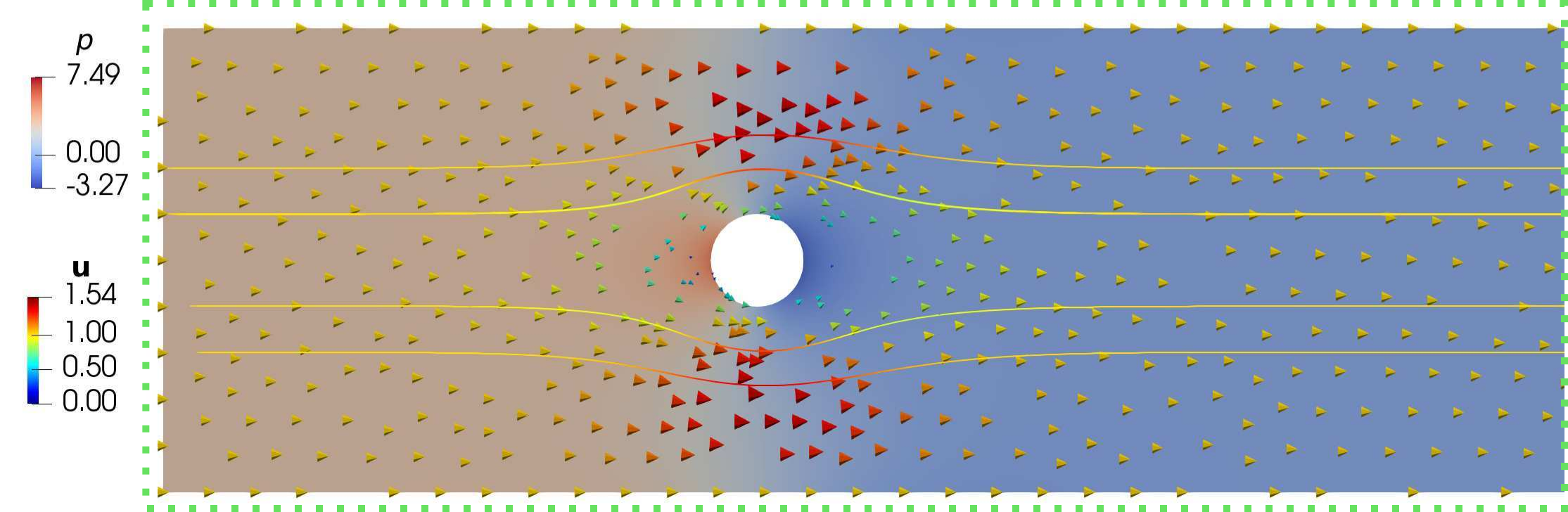} 
\caption{$\beta=1$}
\end{subfigure}
\begin{subfigure}{0.9\textwidth}
\includegraphics[width=0.7\textwidth]{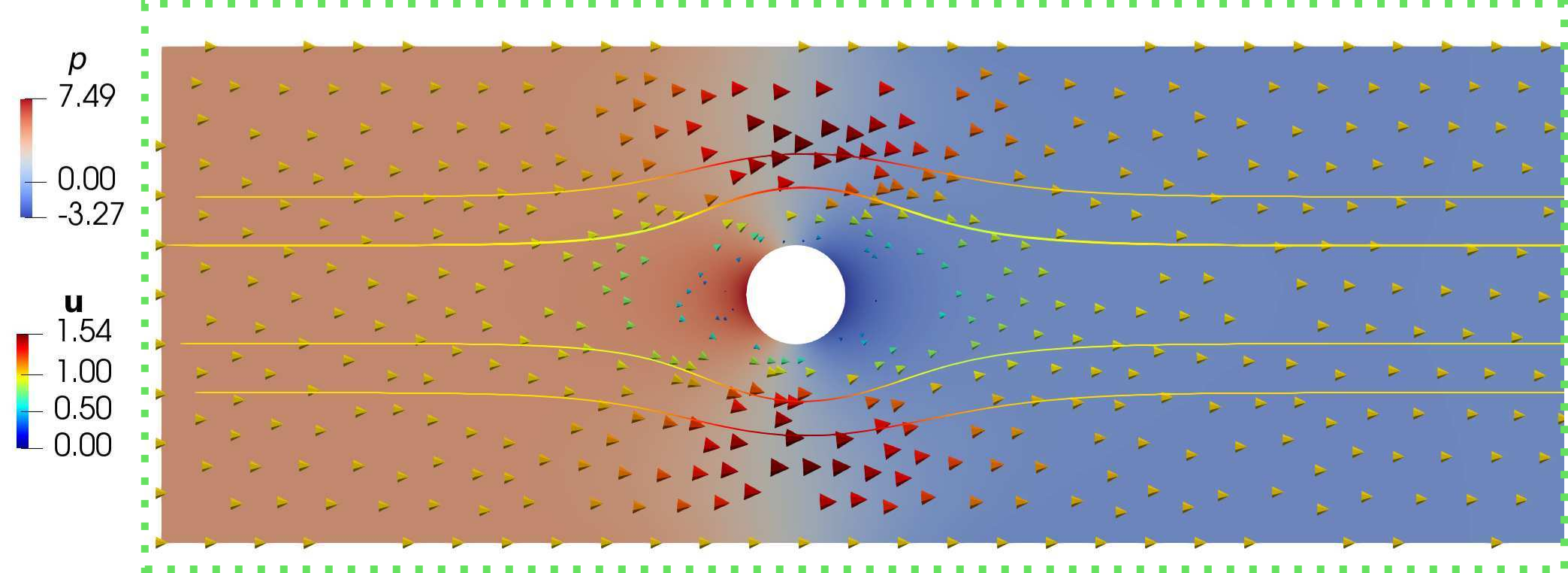} 
\caption{$\beta=100$}
\end{subfigure}
\caption{Closeup views of the pressure and velocity for $R=1$ and different values of $\beta$. For this small Reynolds number there is an evident difference in the tangential flow velocity along $\Gamma$. This further affects the pressure drop from fore to aft of the cylinder, and explains why these flows have different pressure drag coefficients.}
\label{fig:closeup}
\end{figure}

In Figure \ref{fig:betaK} we see the pressure and velocities for $\beta=100$ and $R\in \{ 1, 10, 100, 1000 \}$. Visually, there is only a small change from $\beta=1$ for the larger Reynolds numbers. The difference is larger for $R=1$. For comparison, Figure \ref{fig:closeup} gives a closeup view for $\nu=1$ and $\beta=1$ (top) and $\beta=100$ (bottom). We see that the tangential flow along $\Gamma$ is substantially larger using the smaller friction parameter $\beta=1$. This leads to a smaller pressure drop for $\beta=1$ compared to for $\beta=100$; consequently, the drag coefficient is also smaller.

The tangential velocity is further examined in Table \ref{table:l2-norm}. There, we give the $L^2$-norm of $\uu$ on $\Gamma$. We see that the magnitude of the tangential velocity significantly depends on $\beta$. Thus, we do expect significant dependence of the viscous drag on $\beta$.

Finally, let us comment further on the values for the viscous drag versus pressure drag. We see from Table \ref{tab:viscdrag} that the viscous drag coefficients are quite small, and decreasing as the viscosity decreases. But the viscous drag coefficient is not simply proportional to $\nu$.
The shear strain is increasing like the square root of the Reynolds number. 
Thus other functionals may not behave well as the Reynolds number increases.
Note that the wake length is also growing as the Reynolds number increases, so it is unclear
that a limiting solution exists in a simple sense.

The viscous/skin-friction drag is multiplied by the parameter $\nu$, so
as $\nu\to 0$, this term may become quite small. For small $\nu$, there is a balance in the equation \eqref{eqn:firstnavst} between $\uu\cdot\nabla\uu$ and $\nabla p$, so the form/pressure drag is closely related 
to the nonlinear term in \eqref{eqn:firstnavst}.

\begin{table}
\begin{center}
\begin{tabular}{|c|*{4}{c|}}
\hline
\backslashbox[2em]{$\beta$}{$R$} & 1 & 10 & 100 & 1000 \\ 
\hline
0  & 1.388 &  1.729 &  2.648 &  3.315 \\ 
1  & 1.198 &  0.727 &  0.292 &  0.099 \\ 
10  & 0.538 &  0.118 &  0.033 &  0.010 \\ 
100  & 0.083 &  0.013 &  0.003 &  0.001 \\ 
\hline
\end{tabular}
\end{center}
\vspace{-6mm}
\caption{$L^2$-norms for $\uu$ restricted to the cylinder boundary; i.e. $\norm{\uu}_{L^2(\Gamma)}$.}
\label{table:l2-norm}
\end{table}

\subsection{Comparison with experimental measurements}
\label{sec:expmeass}

\begin{figure}
    \centering
    \includegraphics[width=0.7\textwidth]{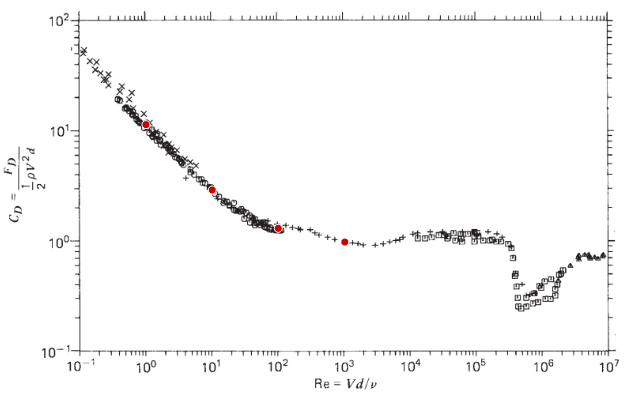}
    \caption{Experimental measurements of the drag coefficient \cite[Figure 14.15]{PantonRonaldL2013IF}. Red points indicate the data points used to estimate the drag coefficients listed in Table \ref{tab:fulldrag} at $R=1,10,100$ and $1000$. The drag value at each of these points was extracted using WebPlotDigitizer \cite{Rohatgi2020} and are listed in Table \ref{tab:fulldrag}.Permission to reproduce \cite[Figure 14.16]{PantonRonaldL2013IF} 
was granted by John Wiley and Sons via License Number 5192730691001.}
    \label{fig:exp-values}
\end{figure}

The drag coefficient has been measured experimentally 
\cite{goldstein1950modern,ref:cylendardralow,ref:cylendardrag}
for various geometries $\Gamma$.
A combination of experimental and computational data for a cylinder is reported 
in \cite[Figure 154]{goldstein1950modern} for $R\geq 10$, up to just less than $10^6$.
In \cite[Figure 149]{goldstein1950modern}, theoretical predictions for smaller
values of $R$ are given.
Experimental values for the pressure drag in the range $10\leq R\leq 10^5$ are given in
\cite[Figure 2b]{ref:pressuredrag}, and \cite[Conclusions]{ref:pressuredrag} they state
that in this range
``the mean pressure drag coefficient was confined within the interval $1.05\pm 0.2$.''

Figure \ref{fig:exp-values} reproduces \cite[Figure 14.15]{PantonRonaldL2013IF} which summarizes
the results of several papers measuring the drag coefficient for flow past a cylinder. The drag coefficient was extracted using WebPlotDigitizer \cite{Rohatgi2020} for $\nu=1, 10, 100$ and 1000. The results are given in the bottom row of Table \ref{tab:fulldrag}. The drag coefficient computed from in our numerical simulations were seen to match closely those measured experimentally.  

Additional experimental data for a cylinder \cite[Figure 1]{ref:cylendardrag}
confirms that $C_D$  is essentially a constant, close to $1$, for $R$ greater than $10^4$ and less than $3\times 10^5$ or so. This behaviour is also seen in Table \ref{tab:fulldrag} for the drag coefficient computed from simulations. For $3\times 10^5\leq R \leq 5\times 10^5$, $C_D$ drops by nearly a factor of 5 \cite[Figure 1]{ref:cylendardrag}, as is also indicated in \cite[Figure 154]{goldstein1950modern} by a dashed line. The same is seen in Figure \ref{fig:exp-values}. This phenomenon is known as the drag crisis. 

The drag crisis \cite{hoffman2006simulation,ref:spherecylinderthesis} occurs 
at about $R=5\times 10^5$, or for a cylinder of diameter 20 centimeters, 
at about 90 kilometers per hour in air. Experiments with Reynolds numbers in this range could be done on the top of a car
on a freeway without breaking the speed limit. The drag crisis is not fully explained, but it has the character of a phase transition. It may be related to a transition to a Beltrami flow \cite{ref:LinkeBeltramiFlows}. Thus it is reasonable to think of flow for Reynolds numbers much larger than this as having a different character. For our numerical simulations, we have managed to compute stable solutions of the Navier--Stokes equations for $R \sim10^3$. It is known that there exists steady-state solutions to the Navier--Stokes equations (at least with Dirichlet boundary conditions) for all Reynold's numbers \cite{THESE_1933__142__1_0, Hopf1950berDA} (see also \cite[Lemma 9.2]{hermint2002}). However, we see that as the Newton continuation method moves to higher Reynold's numbers (i) the computational complexity increases as the mesh size is reduced by the adaptive solver to account for the cylinder wake and (ii) the Newton continuation requires smaller and smaller steps. Thus, the solution gets prohibitively expensive to compute.

\subsection{Friction scaling}
\label{sec:frikskal}

The friction coefficient scaling can also be determined.
In dimensional coordinates, \eqref{eqn:navierslip} becomes
\begin{equation} \label{eqn:dimenrslip} 
\hat\beta \,\hat \uu\cdot\hat\btau_k = -\hat\nu \, 
         \hat\nn^t(\nabla\hat\uu+\nabla\hat\uu^t)\hat\btau_k,\quad k=1,2.
\end{equation}
Using the same change of variables as before, we get
$$
\hat\beta U \,\uu\cdot\btau_k = -\hat\nu \, L^{-2} \nn^t(\nabla\uu+\nabla\uu^t)\btau_k,\quad k=1,2.
$$
Multiplying by $L$ and dividing by $U$, we get
\begin{equation} \label{eqn:dimenagain} 
\hat\beta L \,\uu\cdot\btau_k = -\nu \, \nn^t(\nabla\uu+\nabla\uu^t)\btau_k,\quad k=1,2.
\end{equation}
Thus $\beta=\hat\beta L$.
This indicates that $\hat\beta$ has dimensions inverse length.
The quantity $1/\hat\beta$ is known as the slip length \cite{jimenez2017derivation}.

Less is known about the friction coefficient $\hat\beta$ for physical materials
(say air and aluminum) than is known about viscosity (see Table \ref{tabl:visconly}).
But some computational experiments have addressed the estimation of the slip length
\cite{ref:slipwettingcontangle,ref:SlipNewtonianReviewexpts,ref:molecularLiquidSlip}.
Slip lengths are estimated to be on the order of hundreds of nanometers.
For an airplane in flight, with wing thickness $L$ of 1 meter, this means that
$\beta\approx 10^{7}$.

At 225 miles per hour ($\approx 100$ meters per second), at minus forty degrees (C or F), 
the Reynolds number would be about $10^7$, with appropriate corrections for altitude.
But our data in Table \ref{tab:fulldrag} indicates that drag does not change much as
$\beta$ is increased.

\begin{table}[h]
\begin{center}
\begin{tabular}[t]{|c|c|c|}
\hline
fluid & $\hat\nu$ &  temperature \\
\hline
air & 0.170 cm$^2$/sec &  40 C = 104 F \\
air & 0.147 cm$^2$/sec &  15 C = 59 F \\
air & 0.100 cm$^2$/sec & $-40$ C =$-40$ F \\
\hline
water & 0.013 cm$^2$/sec &  10 C = 50 F  \\
water & 0.010 cm$^2$/sec &  20 C = 68 F  \\
water & 0.006 cm$^2$/sec &  45 C = 113 F \\
\hline
\end{tabular}
\end{center}
\vspace{-6mm}
\caption{Kinematic viscosity coefficients $\hat\nu$ for air and water \cite{ref:waterviscosity} 
in units of cm$^2$/second.}
\label{tabl:visconly}
\end{table}

\section{Impact of slip on Newton continuation method}
\label{sec:slipimact}

In the previous section, we showed that for $\beta$ large enough the drag coefficients we compute are approximately the same as computed using a no-slip condition as reference. This raises the question of why it is preferable to work with the Navier-slip condition using large $\beta$ rather than imposing no-slip directly. 

It is known that the slip condition eliminates boundary layers \cite{ref:slipNStoEuler,ref:slipNStoEulerbis,ref:slipNStoEuler3D,ref:variableslipNStoEuler}. 
It also has significant computational advantages.
The Navier slip boundary condition can be thought of in optimization parlance as a relaxation
of the hard constraint of a Dirichlet boundary condition.
This relaxation appears to have little impact on the drag coefficient but is more
efficient computationally.

\begin{figure}
\centering
\includegraphics[width=0.6\textwidth]{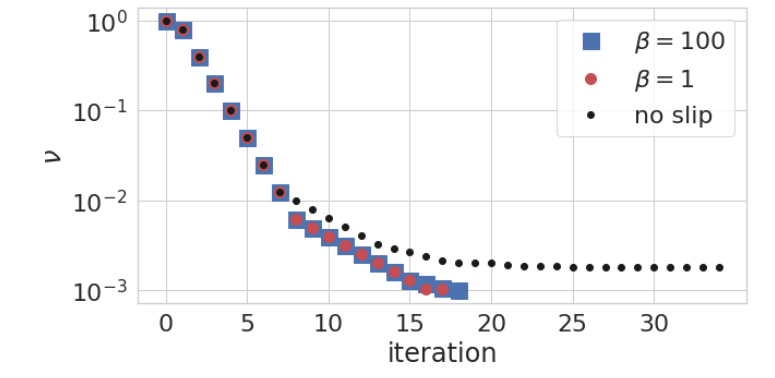}
\caption{Newton continuation method iterations with $\nu_\text{start}=1$ and target value $\nu=10^{-3}$. The solver uses either the slip boundary condition ($\beta=100$: blue square markers and $\beta=1$ round red markers) or a no-slip boundary condition (black dot markers). The continuation method only succeeds using slip boundary conditions. For the no-slip boundary condition, the continuation method encounters a bifurcation at $\nu\sim 1.8\times 10^{-3}$.}
\label{fig:newton}
\end{figure}

The Newton solver we use to solve the non-linear system \eqref{eqn:nitsche} requires a good initial guess in order to converge, especially for high Reynolds numbers. This is handled by the Newton continuation method \cite[section 9.4]{lrsBIBih}. The Newton continuation was implemented using the following methodology: First, the Navier--Stokes equations are solved using an adaptive linear variational solver for an initial $\nu_\text{start}$. For this first step, the Newton solver is given $\uu=(0,0,0)$ and $p=0$ for the initial guess. The Newton solver is given a maximum amount of iterations it is allowed to perform in order to get the error less than some given tolerance. In order for the Newton iterations to converge, the initial viscosity $\nu_{\text{start}}$ should be chosen high enough that the Reynold's number is still small. Upon successfully solving the Navier--Stokes equations for $\nu_{\text{start}}$, the algorithm proceeds to solve for $\nu=0.5 \nu_{\text{start}}$ using the previously computed solution as the initial guess for the Newton solver. If this fails, the algorithm tries to solve for $\nu=0.8 \nu_{\text{start}}$. Should this fail, the algorithm tries to solve for $\nu=0.9 \nu_{\text{start}}$, and so on. Thus the distance between the initial guess and the solution is decreased every time the solver fails. The precise implementation of this method is indicated in Listing \ref{lst:cont}. The implementation of the adaptive linear variational solver is indicated in Listing \ref{lst:solver}.

In Figure \ref{fig:newton}, we show the iterations made by a Newton continuation method iterating from low to large Reynolds numbers. The continuation method was started with $\nu_\text{start}=1$ and the target value was set to $\nu=1 \times 10^{-3}$. We see that the continuation method is more efficient for the Navier slip boundary condition than for the (Dirichlet-type) no-slip boundary condition. The continuation method using the no-slip boundary condition appears to encounter a bifurcation at $\nu\sim 1.8 \times 10^{-3}$, where the Newton iteration fails to converge even as the continuation method is making very small steps towards the higher Reynold's number. Using the Navier slip boundary condition we see that the Newton continuation was able to reach the target value of $\nu=1\times 10^{-3}$ in less than 20 iterations.

\section{Comparison with asymptotic expansions}
\label{sec:asymptions}

Asymptotic expansions \cite[Chapter 15]{PantonRonaldL2013IF} write solutions 
in a series involving a small parameter to different powers for the separate terms.
For example, the hydrogen molecule H$_2$ is studied in \cite{lrsBIBhi} where the small 
parameter is one over the separation distance between the two hydrogens.
The key is that the equations for the individual terms in the expansion are simpler,
but they typically depend on lower-order terms in the expansion.
In the case of H$_2$, the wave function for infinitely separated atoms is known analytically.
This allows computation of the next term in the expansion, which provides an accurate 
prediction of the separation energy for large separation distances.

Matched asymptotic expansions involve two such expansions, one that might be valid 
in one part of the domain and the other in another part of the domain.
The equations for each term may involve lower-order terms in both expansions, and free parameters
are often determined by matching across the interface of the two domains.
The Triple Deck involves as well an intermediate domain \cite{stewartson1981d}.
We refer to \cite{stewartson1981d} and Wikipedia for more information regarding this approach.

We have indicated that the viscosity may not be appropriate as a small parameter.
Indeed, we see that the drag on a cylinder is not really constant, nor does it seem to converge
to an obvious limit as the Reynolds number increases.
Moreover, the values of interest occur at finite ranges of viscosity values (Reynolds numbers)
where the behavior can be complex, especially the drag crisis.
However, we have demonstrated that the drag coefficients can be computed for steady flows at least
for moderately large Reynolds numbers, and the results are in good agreement with experiment.

\section{Conclusions}

We show that using a sufficiently positive friction coefficient in Navier's slip boundary condition
gives a drag coefficient that is in agreement with experimental values,
resolving d'Alembert's Paradox.
Moreover, once the friction coefficient is sufficiently positive, the drag is
largely independent of the coefficient.
And these values agree with what is obtained with the Dirichlet (Stokes no-slip) condition, the 
formal limit of the friction boundary condition for large coefficients.
Prandtl \cite{ref:PrandtlresolveDalembrt} proposed resolving d'Alembert's paradox
using the Stokes no-slip boundary condition together with a boundary layer
theory that he developed.
It is known that the slip condition eliminates boundary layers
\cite{ref:slipNStoEuler,ref:slipNStoEulerbis,ref:slipNStoEuler3D,ref:variableslipNStoEuler},
which is consistent with the numerical simulations presented here.

We showed that the flow strain appears to go to infinity like the square root
of the Reynolds number and that the wake (recirculation) region behind the cylinder
also grows substantially in length as the Reynolds number is increased.
These two observations suggest caution regarding theoretical studies predicated
on a limit as the viscosity goes to zero.


\section{Acknowledgments}

Permission to reproduce \cite[Figure 14.16]{PantonRonaldL2013IF} 
was granted by John Wiley and Sons via License Number 5192730691001 and is gratefully acknowledged.

\bibliographystyle{plain}

\appendix
\newpage
\section{Implementation}

The Navier--Stokes equation with slip boundary condition \eqref{eqn:firstnavst}-\eqref{eqn:navierslip} were solved using the Nitsche method \eqref{eqn:nitsche} to weakly impose boundary conditions. The system was discretized using continuous piecewise quadratic $\mathbb{P}^2$ functions for the velocity and continuous piecewise linear $\mathbb{P}^1$ functions for the pressure. The initial mesh was generated using GMSH \cite{GMSH}, with refined mesh cells around $\Gamma$. The resulting system was solved in FEniCS \cite{fenics2012} using a adaptive linear variational solver, with the goal functional $M = \int_\Omega \vert \nabla \mathbf{u} \vert^2 \, \mathrm{d}x$ and tolerance 0.25. The adaptive solver was used to account for the fact that the wake behind the cylinder grows as the Reynold's number increases. I.e., the wake gets sharper as the Reynold's number increases, leading to larger values for $\nabla \uu$ at the boundary of the wake. Thus the mesh needs to be refined in this area in order to accurately capture the wake.

\begin{lstlisting}[basicstyle=\scriptsize, language=Python,label={lst:cont}, caption= Implementation of Newton continuation method.]
def newton_continuation_method(nu_start, nu_target, mesh, beta, gamma):
    
    nu_prev = nu_start
    
    # At each iteration of the continuation method we attempt to solve the Navier-Stokes equations with a smaller value for nu, giving our previous solution as the initial guess to the Newton solver
    
    # For the first iteration we set the initial guess to zero
    up_prev = Constant((0.0, 0.0, 0.0, 0.0))
    
    # nu is reduced according to the rule nu = cont_factor*nu_prev
    # where cont_factor comes from a list of factors we try
    continuation_factors = [0.5, 0.8, 0.9, 0.95, 0.999, 0.9999]
    
    # We keep track of how many times the continuation method fails
    failure_tracker = 0
    cont_factor = continuation_factors[failure_tracker]
    
    # If the Newton solver does not converge, we increment the failure tracker and update cont_factor, thus trying with a value for nu closer to the one we previously computed
    
    while nu_prev > target_nu:

        try:
            nu = nu_prev*newton_cont_factors[failure_tracker]
            
            # Attempt to solve the Navier Stokes equations for this value of nu
            up_sol = solver(mesh, beta, gamma, nu, up_prev)
            
            # Save solutions and proceed to next continuation iteration
            save_solution(up_sol, beta, nu)
            
            up_prev = up_sol
            nu_prev = nu
            
            failure_tracker -= 1
            
        except:
        
            # If the Newton solver does not converge in a certain amount of steps, the solver throws an exception
            
            failure_tracker += 1  
    
\end{lstlisting}

\begin{lstlisting}[basicstyle=\scriptsize, language=Python, label={lst:solver}, caption= Implementation of adapative solver for Navier--Stokes equations with slip boundary conditions.]
from fenics import *

def solve(mesh, beta, gamma, nu, u_prev, p_prev)
    
    # Make FE space with Taylor-Hood elements with given degree
    P2 = VectorElement("Lagrange", mesh.ufl_cell(), degree=2)
    P1 = FiniteElement("Lagrange", mesh.ufl_cell(), degree=1)
    LM = FiniteElement('R', mesh.ufl_cell(), 0)

    TH = MixedElement([P2, P1, LM])
    W = FunctionSpace(mesh, TH)

    # We use the "projected normal" (by taking the normal vector from the continuous flow domain)
    n = Expression(('-x[0]', '-x[1]'), degree=2)
    tau = as_vector([n[1], -n[0]])

    # Define variational problem
    (u, p, rho) = split(up)
    (v, q, lamda) = TestFunctions(W)
    
    h = mesh.hmin()

    ds = Measure("ds", domain=mesh, subdomain_data=boundary_markers, subdomain_id=1)
    dx = Measure('dx', domain=mesh)

    a =  Constant(0.5*nu)*inner(D(u), D(v))*dx
    a += Constant(beta) * dot(u, tau) * dot(v, tau) * ds
    a += - Constant(nu)*dot(D(u)*dot(v,n)*n, n)*ds
    a += - Constant(nu)*dot(D(v)*dot(u,n)*n, n)*ds
    a += Constant(gamma) * Constant(1.0 / h) * dot(u, n) * dot(v, n) * ds
    a += inner(grad(u)*u, v)*dx - div(v)*p*dx - q*div(u)*dx
    a += rho*q*dx + lamda*p*dx
    a += p*dot(v,n)*ds + q*dot(u,n)*ds
 
    L =  dot(Constant((0.0, 0.0)),v)*dx

    # On the sides of the box, we use an essential boundary condition for velocity
    bc = DirichletBC(W.sub(0), Constant((1.0, 0.0)), boundary_markers, marker_value)

    # Define adaptive variational problem
    M = inner(grad(u), grad(u))*dx
    F = a - L 
    JF = derivative(F, up, TrialFunction(W))
    problem = NonlinearVariationalProblem(F, up, bc, JF)
    parameters["refinement_algorithm"] = "plaza_with_parent_facets"

    solver = AdaptiveNonlinearVariationalSolver(problem, M)

    prm = solver.parameters["nonlinear_variational_solver"]
    prm['newton_solver']['absolute_tolerance'] = 5E-8
    prm['newton_solver']['maximum_iterations'] = 5
    
    # Compute solutions to within given tolerance
    tol  = 0.25
    solver.solve(tol)
    return up
\end{lstlisting}

\end{document}

%% file: machip.tex
\newcommand{\btau}{{\boldsymbol \tau}}
\newcommand{\omitit}[1]{}

\newcommand{\p}{{{\rm I}{\relax\ifmmode\mskip-\thinmuskip\relax\else\kern-.22em\fi}{\bf P}}}
\newcommand{\perpzh}{\perp\kern-.20em^a}

\newcommand{\Plang}{{\p}{\kern-.03em{\rm fortran}}}
\newcommand{\Pfortran}{{\p}{\kern-.03em{\rm fortran}}}

\newcommand{\half}{{\textstyle{1\over 2}}}

\newcommand{\set}[2]{\left\lbrace #1 \; : \; #2 \right\rbrace}

\newcommand{\bo}{{\partial\Omega}}

\newcommand{\du}{{\cal D}(\uu)}

\newcommand{\dhu}{{\cal D}(\hat\uu)}
\newcommand{\dz}{{\cal D}(\zz)}

\newcommand{\dv}{{\cal D}(\vv)}

\newcommand{\dvl}{{\cal D}(\vv_\lambda)}
\newcommand{\dchi}{{\cal D}(\bchi)}

\newfloat{code}{ht}{aux}[section]
\floatstyle{boxed}
\floatname{code}{Program}
\restylefloat{code}

\newcommand{\norm}[1]{\Vert{#1}\Vert}

\newcommand{\tbnorm}[1]{\vert\kern-.1em\vert\kern-.1em\vert{\, #1 \,}\vert\kern-.1em\vert\kern-.1em\vert}

\newcommand{\oput}[1]{\rlap{${}{\lower 1.3ex\hbox{${\sim}$}}$}{#1}}

\newcommand{\sdiv}{{{\nabla\cdot} \,}}

\newcommand{\bfz}{{\mathbf 0}}

\newcommand{\uu}{{\mathbf u}}

\newcommand{\vv}{{\mathbf v}}
\newcommand{\ww}{{\mathbf w}}
\newcommand{\zz}{{\mathbf z}}
\newcommand{\gbc}{{\mathbf g}}

\newcommand{\nn}{{\mathbf n}}
\newcommand{\ff}{{\mathbf f}}
\newcommand{\xx}{{\mathbf x}}

\newcommand{\intox}[1]{\int_{\Omega} #1 \, d\xx}

\newcommand{\eps}{{\epsilon}}

\newcommand{\bchi}{{\boldsymbol \chi}}

\newcommand{\datekern}{{\relax\kern+.1em}--{\relax\kern-.01em}}

\newcommand{\antiparallel}{\not\kern-0.35em\Vert}